# Simulation of longitudinal dynamics of laser-cooled and RF-bunched $C^{3+}$ ion beams at heavy ion storage ring CSRe*

Xiao-Ni Li (李小妮)[1,2], Wei-Qiang Wen (汶伟强)[1], Heng Du (杜衡)[1,2], Peng Li (李朋)[1], Xiao-Hu Zhang (张小虎)[1], Xue-Jing Hu (胡雪静)[1], Guo-Feng Qu (曲国锋)[1,2], Zhong-Shan Li (李钟汕)[1,2], Wen-Wen Ge (葛文文)[1,2], Jie Li (李杰)[1], Han-Bing Wang (汪寒冰)[1,2], Jia-Wen Xia (夏佳文)[1], Jian-Cheng Yang (杨建成)[1], Xin-Wen Ma (马新文)[1], You-Jin Yuan (原有进)[1; 1)]

[1] Institute of Modern Physics, Chinese Academy of Sciences, 730000, Lanzhou, China
[2] University of Chinese Academy of Sciences, 100049, Beijing, China

**Abstract:** Laser cooling of relativistic heavy ion beams of Li-like $C^{3+}$ and $O^{4+}$ is being in preparation at the experimental Cooler Storage Ring (CSRe). Recently, a preparatory experiment to test important prerequisites for laser cooling of relativistic $^{12}C^{3+}$ ion beams using a pulsed laser system has been performed at the CSRe. Unfortunately, the interaction between the ions and the pulsed laser cannot be detected. In order to study the laser cooling process and find the optimized parameters for future laser cooling experiment, a multi-particle tracking method was developed to simulate the detailed longitudinal dynamics of laser-cooled ion beams at the CSRe. The simulations of laser cooling of the $^{12}C^{3+}$ ion beams by scanning the frequency of RF-buncher or continuous wave (CW) laser wavelength were performed. The simulation results indicated that a large momentum spread ion beams could be laser-cooled by the combination of only one CW laser and the RF-buncher and shown the requirements of a successful laser cooling experiment. The optimized parameters for scanning the RF-buncher frequency or laser frequency were obtained. Furthermore, the heating effects were estimated for laser cooling at the CSRe. The Schottky noise spectra of longitudinally modulated and laser-cooled ion beams was simulated to fully explain and anticipate the experimental results. The combination of Schottky spectra from the highly sensitive resonant Schottky pick-up and the simulation methods developed in this paper will be helpful to investigate the longitudinal dynamics of RF-bunched and ultra-cold ion beams in the upcoming laser cooling experiments at the CSRe.

* Supported by the National Natural Science Foundation of China (11405237, 11504388)
1) yuanyj@impcas.ac.cn

## 1. Introduction

Laser cooling has been considered as one of the most powerful techniques to reach high phase space densities for the relativistic heavy ion beams at the storage rings [1, 2]. Compared with the stochastic cooling [3, 4] and electron cooling [5, 6], the laser cooling is expected to obtain much higher cooling rate, which results in higher phase space densities to achieve phase transition for ordered beam or crystallized beam [7-9]. So far, beams of several ion species have been cooled at heavy ion storage rings successfully, such as $^7Li^+$ ion beams at the TSR in Heidelberg and the ASTRID in Aarhus, $^{24}Mg^+$ ion beams at the S-LSR in Kyoto, and $^{12}C^{3+}$ ion beams at the ESR in Darmstadt [10-16]. However, none of the laser cooling experiments has achieved phase-transition to obtain crystallized highly charged heavy ion beams, there are still many challenges to reach this states. The systematic simulation and the experimental study are urgent and significant to be performed for laser cooling progress at the storage ring. Currently, the much of laser cooling simulation and experiments are concentrate on the longitudinal phase space. The transverse laser cooling of stored ion beams can be realized through the intra beam scattering (IBS) effect and dispersion coupling effect which were investigated at the TSR [17, 18] at MPIK and the S-LSR at Kyoto University [19]. However, it will be very difficult to investigate transverse laser cooling of relativistic ion beams at the CSRe.

Laser cooling experiments of relativistic $^{12}C^{3+}$ and $^{16}O^{4+}$ ion beams are making progress at the experimental Cooler Storage Ring (CSRe) in Lanzhou [20-22]. Recently, a test experiment towards laser cooling of $^{12}C^{3+}$ ion beams at an energy of 122 MeV/u was performed with a pulsed laser system. The aims of the experiment were to test the experimental method of building laser system in accelerator machine and overlapping the laser with ion beams. For laser cooling schemes, the CW laser is used for bunched ion beams. In order to investigate the dynamics of the laser-cooled and RF-bunched heavy ion beams and the properties of ultra-cold ion beams, the molecular dynamics simulation was carried out at the storage ring S-LSR, which is a compact ion storage ring in Kyoto University [2]. A multi-particle tracking simulation method was developed in this paper. The longitudinal dynamics of the RF-bunched and laser-cooled $^{12}C^{3+}$ ion beams at the kinetic energy of 122 MeV/u were simulated to study the laser cooling process and find the optimized parameters of the cooling process at the CSRe. The heating effects were simulated. Moreover, a simulation to produce the Schottky noise spectra of the RF-bunched and laser-cooled ion beams was performed to explain and anticipate the experimental results, especially for ultra-cold ion beams. The combination of the simulation and the experiments will be very helpful to investigate the properties of ultra-cold ion beams in the future at the CSRe.

In the paper, a brief description of the setup for laser cooling experiment at the CSRe, and the details of the simulation method and the simulation results of laser cooling will be presented. Last but not the least, analysis of the Schottky noise spectra of the RF-bunched and laser-cooled ion beams will be introduced and discussed.

## 2. Experimental setup for laser cooling of relativistic $C^{3+}$ ion beams at the CSRe

The layout of laser cooling setup at the CSRe is shown in Fig. 1. The laser cooling experiment of relativistic heavy ion beams at the CSRe has already been described in other paper [23]. For laser cooling experiments, $^{12}C^{3+}$ ions were produced, accelerated, and finally injected into the CSRe at an energy of 122 MeV/u (at a velocity of 47% of the speed of light). A suitable laser overlaps the ion beams all along the straight section (~ 20 m) where the new Schottky-noise pick-up and optical diagnostic of UV-PMT and CPM are located. The UV-PMT and CPM are the ultraviolet sensitive photomultiplier tube and Channeltron photomultiplier to detect the fluorescence. A special designed RF-buncher system has been installed to bunch the beam [21]. The relevant parameters for laser cooling of $C^{3+}$ ion beams at the CSRe are listed in table 1.

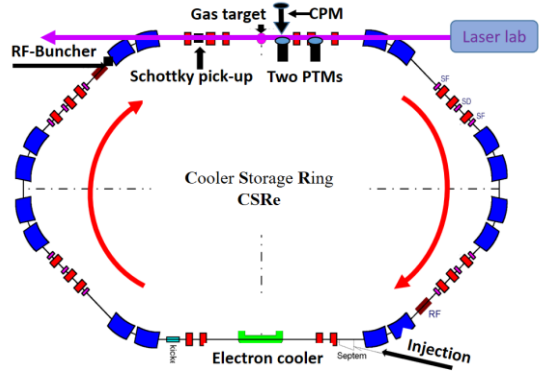

Fig. 1. (color online) Schematic of the CSRe and the experimental setup for laser cooling.

Table 1. The main parameters for laser cooling experiments at the CSRe

| CSRe parameters | |
| --- | --- |
| Circumference | 128.80 m |
| Ion species | $^{12}C^{3+}$ |
| Beam energy | 122 MeV/u |
| Relativistic factor $\beta\ \gamma$ | 0.47, 1.13 |
| Revolution frequency | 1.087 MHz |
| Transition energy $\gamma_t$ | 2.629 |
| Slip factor $\eta$ | 0.64 |
| Harmonic number | 15 |
| Laser system | |
| Laser source | Ar |
| Laser type | CW laser |
| Laser wavelength | 257.5 nm |
| Laser power | 40 mW |
| Scan range | 26 GHz/10ms |
| Cooling optical transition | |
| Optical transition $2S_{1/2} \rightarrow 2P_{1/2}$ | 155.07 nm |
| Natural line width $\Gamma$ | 42 MHz |
| RF-buncher system | |
| Voltage | 1V |
| Harmonic number | 15th |
| Scan range | ~500Hz |

## 3. Multi-particle tracking simulation method of RF-bunched and laser-cooled ion beams

Laser cooling of the relativistic heavy ion beams at the storage ring CSRe will be achieved by the combination of a CW laser and a RF-buncher, as described in reference [21]. The purposes of the RF-buncher include bunching the ion beams, restricting the ions in the bucket and providing a counter-force for laser scattering force. Ion beams with large momentum spread could be laser-cooled by scanning the frequency of the RF-buncher to bring the ions to resonantly interaction with the fixed laser. In laser cooling simulation, the potential of the RF-buncher is sinusoidal wave, as:

$$V = V_o \sin 2\pi h \phi \qquad (1)$$

Here $V_0$ is the voltage amplitude, $h$ is the harmonic number, $\phi$ is the phase value which is equivalent to time. The RF-buncher can be operated at various harmonic number of the revolution frequency. As a result, the ion beams will be modulated longitudinally by the RF-buncher.

In order to study the movement of the RF-bunched $C^{3+}$ ions in longitudinal phase

space, a multi-particle tracking code is developed. The discrete longitudinal dynamics equation is used:

$$\begin{cases} \delta_{n+1} = \delta_n + \dfrac{qeV}{\beta^2 E}\left(\sin\phi_n - \sin\phi_s\right) \\ \phi_{n+1} = \phi_n + 2\pi h\eta\delta_{n+1} \end{cases} \quad (2)$$

Where $\beta$ is the relativistic factor, $qe$ is the ion charge, $\eta$ is the slip factor, $h$ is the harmonic number, $\delta_n$ and $\phi_n$ are the momentum spread and the phase value at nth turn respectively, $\delta_{n+1}$ and $\phi_{n+1}$ are the momentum spread and the phase value at (n+1)th turn respectively. The phase $\phi_{n+1}$ depends on the new off-momentum coordinate $\delta_{n+1}$. The $\phi_s$ is the phase of synchrontron particle [24].

In laser cooling experiment at the CSRe, a CW laser will be employed to resonantly interact with the ion beams at the counter-propagating direction, The optical transition $2s_{1/2} \rightarrow 2p_{1/2}$ of $C^{3+}$ ion with energy of 122 MeV/u, while the transition energy is equivalent to a laser wavelength $\lambda_0$ =155.07 nm could be excited by the laser with wavelength of 257.5 nm, as results of relativistic Doppler effect [25, 26]. However, the longitudinal modulation of the ions will make the ions oscillate inside the bucket of the RF-buncher. In the simplified model, the simulation contains only longitudinal dynamics assuming that the transverse motion is unaffected by the laser. We assume that the laser is always switched on while the ions circulate at the CSRe.

For laser cooling with the CW laser, the spectral linewidth of the laser is very narrow (<10 MHz). As a result, the momentum acceptance of the cooling force is also narrow (~ $5\times10^{-8}$) accordingly. In order to increase the momentum acceptance of the cooling force, it is necessary to scan the frequency of RF-buncher or laser [14, 27].

## 4. The simulation results and discussion

The schemes of laser cooling of RF-bunched ion beams are shown in Fig.2. Two schemes can be employed, (a) scanning RF-buncher frequency while the CW laser frequency is fixed, as a result, the ions inside the bucket will be accelerated to resonantly interact with the laser and then be cooled down and (b) scanning laser frequency while the RF-buncher frequency is fixed, the ions will be pushed into the center of the bucket continually, and then be cooled down. It needs to be noted that, all the main simulation parameters are the same as the experimental one in the multi-particle tracking simulation. Usually, the laser cooling is carry out after electron cooling. For all of the simulation results presents below, the initial momentum spread of the ion beams was set as to ~$1.6\times10^{-5}$ for 3 sigma, which was achieved by the beam parameters after the electron cooling experiment, and a number of 10000 particles with Gaussian distributions was stored in every bunch.

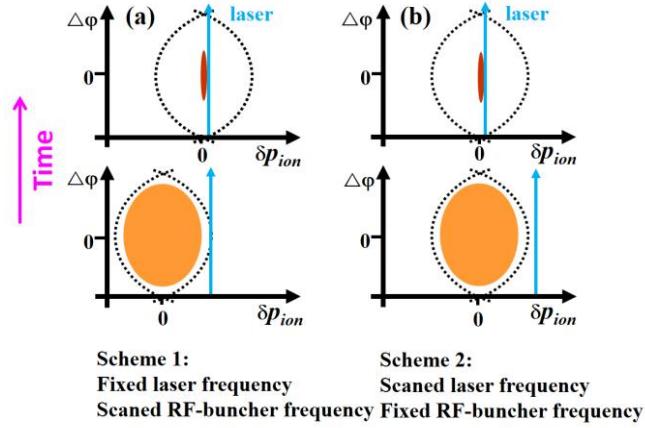

Fig. 2. (color online) Two schemes of laser cooling of relativistic ion beams by a CW laser and a RF-buncher.

### 4.1 Laser cooling with the scanned RF-buncher and fixed laser frequency

Fig. 3 schematically illustrates the principle of laser cooling of RF-bunched ion beams for the first scheme as shown in Fig 2 (a), scanning RF-buncher frequency while the laser frequency is fixed. The phase space trajectory of single particle $C^{3+}$ is shown, and each point represents the phase and energy deviations of the ion in the bucket of RF-buncher. Fig. 4 shows the simulated longitudinal momentum spread as a function of turn numbers for the various scanning speed of the RF-buncher frequency. It can be found that the best scanning speed of the RF-buncher frequency is 20 Hz/s, and the minimum momentum spread approximate $5\times10^{-9}$ can be reached if the heating effects such as intra-beam scattering effect and the space charge effect did not taken into account. However, the ion beams could be heated by the laser scattering force when the RF-buncher frequency scanned excessively. Fig. 5 shows the particles distribution in longitudinal phase space at 0, $2\times10^6$ and $4.4\times10^6$ turns. Hence, by detuning the bunching frequency relative to the laser, the momentum acceptance range of the laser could be increased to cover the initial momentum spread of the ion beams.

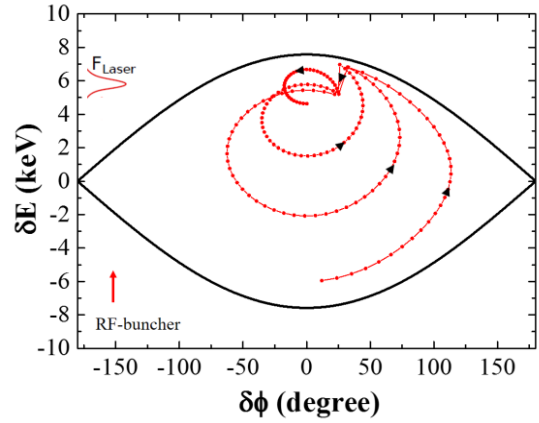

Fig. 3. (color online) Simulation of longitudinal dynamics of laser-cooled single particle $C^{3+}$ in a bunch. Every 2000$^{th}$ turn is indicated by a dot. For the purposes of illustration, the actual cooling rate has been exaggerated by a factor of 1000.

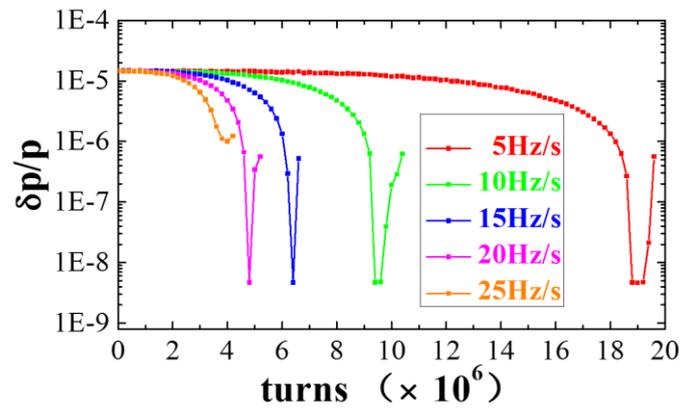

Fig. 4. (color online) Momentum spread reduce versus the turn number during laser cooling. The laser frequency is fixed. Five scanning speeds of the RF-buncher frequency are simulated.

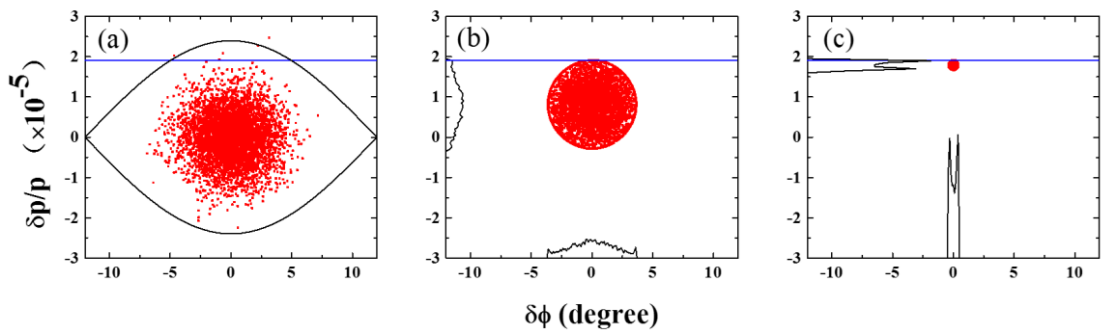

Fig. 5. (color online) The longitudinal phase space distribution of $C^{3+}$ during laser cooling at the different turn, (a) initial distribution at 0 turn, (b) at $2\times10^6$ turns, (c) at

$4.4\times10^6$ turns.

## 4.2 Laser cooling with the scanned laser frequency and fixed RF-buncher frequency

For the laser cooling scheme as shown in Fig. 2 (b), the simulated cooling process of single particle $C^{3+}$ is shown schematically in Fig. 6. The laser frequency is scanned from the separatrix to the center of the bucket, as a result, a large momentum spread ion beams can be cooled down. The various scanning speed of the laser frequency are simulated and shown in Fig. 7. The scanning speed of 0.25 GHz/s for laser is the optimum condition. However, the ion beams will be heated if the laser frequency scanned over the center of the bucket. Fig. 8 shows the particles distribution in longitudinal phase space at 0, $1\times10^7$ and $1.9\times10^7$ turns.

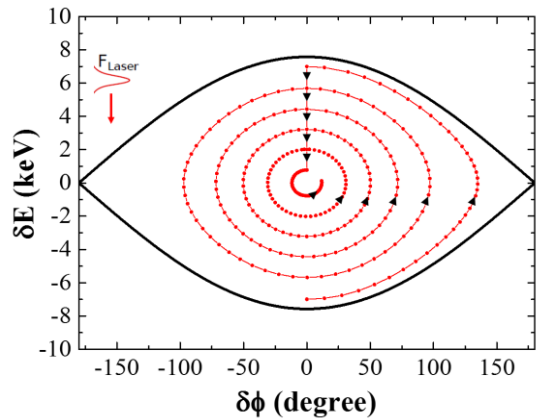

Fig. 6. (color online) Simulation of the cooling process for single particle $C^{3+}$ while the laser frequency is scanned from the separatrix to the center of the bucket. Every $2000^{th}$ turn is indicated by a dot. For the purposes of illustration, the actual cooling rate has been exaggerated by a factor 1000.

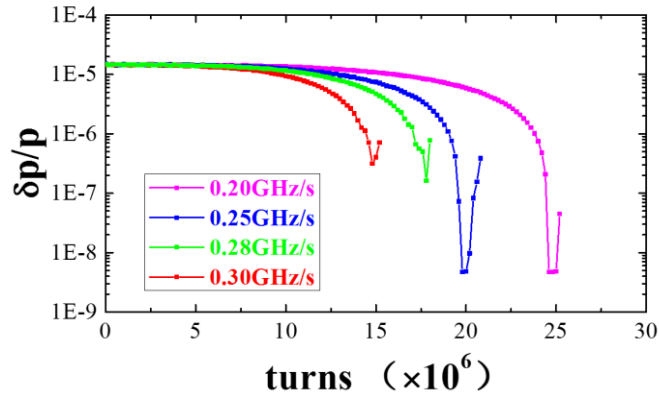

Fig. 7. (color online) The momentum spread reduce versus the turn numbers. Four speeds for scanning the laser frequency are simulated.

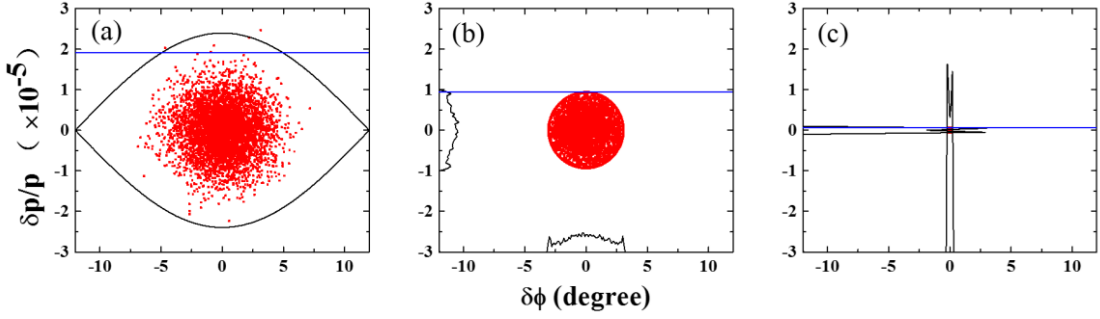

Fig. 8. (color online) The longitudinal phase space distribution of $C^{3+}$ during laser cooling, (a) The initial distribution at 0 turn, (b) at $1\times 10^7$ turns, (c) $1.9\times 10^7$ turns.

### 4.3 Heating effects for laser cooling at the CSRe

In the above simulation results of laser cooling, it is also shown that the laser cooling can obtain much higher cooling rate and very cold ion beams. Therefore, the heating effects will be obvious compared with the stochastic cooling and electron cooling. For the laser-cooled ion beams, the space charge effect and the IBS effect become important. The space charge effect is complicated in the storage ring, it can locally change the distribution of particles in the phase space, it will only become very important for laser cooling experiment when the density of the ions become high. In the simulation, we will not have large number of ions stored inside the CSRe, and the density of the ions inside the bucket will not be so high. In addition, the $C^{3+}$ ions have low charge state and the space charge effect can be carefully ignored in this simulation. The IBS is complicated in the storage ring. It is the key factor which determines the final momentum spread of the laser-cooled ion beams. In the paper, a standard and simplified IBS formulas can be used to calculate the IBS growth rate [28]. It has conformed in the electron cooling experiment and can be written as:

$$\tau^{-1} = \frac{1}{\sigma_p^2}\frac{d\sigma_p^2}{dt} = \frac{r^2 c N \Lambda}{8\beta^3 \gamma^3 \varepsilon_x^{3/2} \langle \beta_\perp^{1/2} \rangle \sqrt{\pi/2}\sigma_s \sigma_p^2} \quad (3)$$

where $r$ is the radius of carbon atom, $c$ is the speed of light in vacuum, $N$ is the particle number, $\Lambda$ is the Coulomb logarithm for IBS ($\Lambda \approx 10-20$), $\beta$ and $\gamma$ are the relativistic factor, $\varepsilon_x$ is the transverse emittance which is a constant in the simulation, $\beta_\perp$ is the average envelope, $\sigma_s$ is the bunch length, $\sigma_p$ is the momentum deviation. For the first scheme of laser cooling, the IBS effect is simulated while the ion beams are cooled by the CW laser, the results are shown in Fig.9. it illustrates the IBS growth increase obviously when the momentum spread is less than $5\times10^{-6}$, the final momentum spread is $2.35\times10^{-6}$ which is a balanced state between the laser cooling and the IBS heating effect.

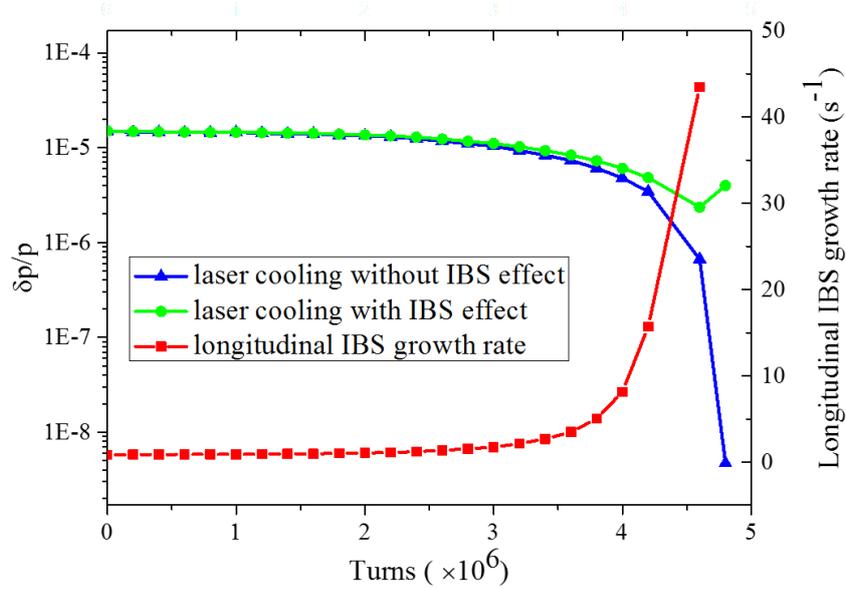

Fig. 9. (color online) Laser cooling with IBS effect, the red carve is the longitudinal IBS growth versus the turn numbers. The blue carve is the momentum spread reduces with the turn numbers without the IBS effect. The green carve is the momentum spread reduces with the turn numbers with IBS effect.

### 4.4 Schottky noise signal of the RF-bunched and laser-cooled ion beams

The longitudinal Schottky signal measurements are a widely used tool for the determination of longitudinal dynamical properties of ion beams at storage ring, such as momentum spread and the synchrotron frequency. In order to compare with the experimental results, a simulation to produce the Schottky noise spectra of the longitudinally modulated and laser-cooled ion beams was performed. For the RF-bunched ion beams, every individual particle executes oscillations at the synchrotron frequency $f_s=\Omega_s/2\pi$. It can be written as

$$f_s = \frac{f_r}{\beta}\sqrt{\frac{qeh\eta U_b}{2\pi\gamma mc^2}} \quad (4)$$

which depends on the revolution frequency $f_r$, the beam velocity $\beta c$, the ion charge $qe$, the momentum slip factor $\eta$, the harmonic number $h$, the effective RF-bunching voltage $U_b$, and the relativistic Lorentz factor $\gamma$, while $m$ is the mass of the ion and $c$ is the speed of light in vacuum. The time of passage of the particle in front of the Schottky pick-up is modulated according to [29]:

$$\tau_i(t) = \hat{\tau}_i \sin(\Omega_s t + \phi_i) \quad (5)$$

$\tau_i(t)$ is the time difference with respect to the synchronous particle (frequency $f_0$) and $\hat{\tau}_i$ is the amplitude of the synchrotron oscillation. In the time domain, the beam current is obtained from the modulated time of passage. It can be written as:

$$i_i(t) = eqf_r + 2eqf_r \, \text{Re}\left\{\sum_{n=1}^{+\infty} \exp jn\omega_r(t + \hat{\tau}_i \sin(\Omega_s t + \phi_i))\right\} \quad (6)$$

Using the relation:

$$\exp[j(z\sin\theta)] = \sum_{p=-\infty}^{+\infty} J_p(z)e^{jp\theta}$$

where $J_p$ is the Bessel function of order $p$. One can expand $n^{th}$ harmonic in equation (6) and obtain:

$$i_n = 2eqf_r \, \text{Re}\left\{\sum_{p=-\infty}^{+\infty} J_p n\omega_r \hat{\tau}_i e^{j(n\omega_r t + p\Omega_s t + p\phi_i)}\right\} \quad (7)$$

each revolution frequency line ($nf_r$) now splits into an infinity of synchrotron satellites, spaced by $f_s$, the amplitudes of which being proportional to the Bessel function of argument $n\omega_r\hat{\tau}_i$ [29].

For the RF-bunched $C^{3+}$ ion beams with the kinetic energy of 122 MeV/u, every individual particle executes synchrotron oscillations at the CSRe. The phase value of the individual ions is $\phi_i$ randomly distributed synchrotron phases from 0 to the total bunch length $\tau_m$. They are collected and equivalent to the time of passage of the ions at the position of Schottky-noise pick-up. It can be written as

$$\phi_i(N) = 2\pi N + \phi_i \quad (8)$$

where $i$ is the particle number, $N$ is the turn number. The particles number at each sampling are counted. The Schottky noise spectrum of the RF-bunched and laser-cooled ion beams in the frequency domain can be obtained from the time domain by a fast Fourier transform (FFT). The simulation result is shown in Fig.10. It shows sharp and pronounced peaks at a spacing determined by the synchrotron oscillation frequency of the ions in the bucket. The synchrotron frequency of the ions inside the bucket can be extracted from the frequency between every two adjacent peaks in the Schottky spectrum. In Fig.10, the abscissa is the frequency deviation relative to the revolution frequency, the center peak represents the revolution frequency, the frequency between every two adjacent peaks represents the synchrotron frequency which is about 326 Hz. It is perfectly accordant with the experimental result [22]. The momentum spread of the RF-bunched ion beams is then given by the synchrotron frequency spread as

$$\frac{\delta p}{p} = \frac{1}{\eta}\frac{\delta f}{hf_r} \quad (9)$$

Which is related to the revolution frequency $f_r$.

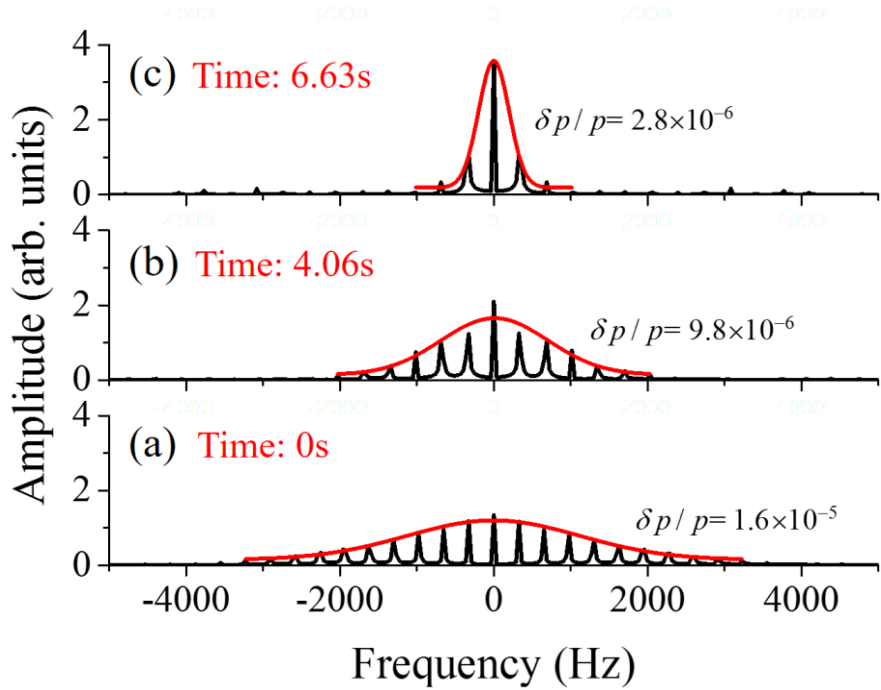

Fig. 10. (color online) The simulation results of the Schottky noise spectrum of the RF-bunched and laser-cooled $^{12}C^{3+}$ ion beams at the time of 0s, 4.06s and 6.63s after injection. The scanning speed of laser frequency is 0.25GHz/s. The momentum spread are $1.6\times10^{-5}$, $9.8\times10^{-6}$ and $2.8\times10^{-6}$ accordingly.

## 5  Conclusions

The longitudinal dynamics of laser cooling of RF-bunched ion beams were simulated by the multi-particle tracking method. The simulation results illustrate that a large momentum spread ion beams could be cooled by the combination of a CW laser and a RF-buncher. The optimized parameters for scanning the RF-buncher frequency or laser frequency are obtained. The IBS effect was estimated at the CSRe, its growth rate increases obviously when the ion beams are very cooled, it determines the final momentum spread of the laser-cooled ion beams. In addition, the Schottky noise spectra of the longitudinal modulated ion beams was produced by the simulation, and the momentum spread of laser-cooled ion beams and the synchrotron frequency of the ions inside the bucket could be deduced. For a successful laser cooling experiments at the CSRe, the frequency of the laser must be appropriate to resonantly interact with the ion beams. The scanning speeds of the RF-buncher frequency and the laser frequency must be suitable. The laser must overlap the ion beams. The combination of simulation and experimental results of laser cooling of relativistic ion beams at the CSRe will provide a novel method to investigate the longitudinal dynamics of ultra-cold ion beams. It needs to be noted that the investigations of laser cooling of heavy ion beams at the CSRe are directly relevant to laser cooling and precision laser spectroscopy of highly charged and relativistic heavy ions at the future large facilities such as High Intensity heavy ion Accelerator Facility (HIAF) in China. At this facility, such experiments could even open up a new field for atomic physics and nuclear physics.

## Acknowledgements

The authors would like to express their sincere thanks to M. Bussmann and L. Eidam for valuable discussion. This work was supported by the National Natural Science Foundation of China (NSFC) through Grant no. 11405237 and 11504388.